\documentclass[%
 aip,
 amsmath,amssymb,
 reprint,%
]{revtex4-1}

\usepackage{graphicx}% Include figure files
\usepackage{dcolumn}% Align table columns on decimal point
\usepackage{bm}% bold math
\usepackage[utf8]{inputenc}
\usepackage[T1]{fontenc}
\usepackage{mathptmx}
\usepackage{etoolbox}
\usepackage[version=3]{mhchem} % Formula subscripts using \ce{}
\usepackage{enumitem}
\usepackage{xcolor}
\usepackage[normalem]{ulem}
\usepackage{graphicx}
\usepackage{wrapfig}
\usepackage{framed}
\usepackage{hyperref} \hypersetup{hidelinks}
\usepackage{color}
\usepackage{float}
\usepackage{physics}

\newcommand{\etal}{\textit{et al.}}
\makeatletter
\def\@email#1#2{%
 \endgroup
 \patchcmd{\titleblock@produce}
  {\frontmatter@RRAPformat}
  {\frontmatter@RRAPformat{\produce@RRAP{*#1\href{mailto:#2}{#2}}}\frontmatter@RRAPformat}
  {}{}
}%
\makeatother
\begin{document}

%\preprint{AIP/123-QED}

\title{Clustering Dynamics of \ce{SiO2}–Pt Active Janus Colloids}
% Force line breaks with \\
 \author{Harishwar Raman}
 \author{Aniket Shivhare}
 \author{Amit Kumar}
 \author{Madhav Penukonda}
 \author{Pawan Kumar}
 \author{Karnika Singh}
 \author{Akash Choudhary} 
 \author{Rahul Mangal*} \email{mangalr@iitk.ac.in}
 \affiliation{Department of Chemical Engineering, Indian Institute of Technology Kanpur, Kanpur 208016 India.}

\date{\today}% It is always \today, today,
             %  but any date may be explicitly specified

\begin{abstract}

Active colloid clustering is central to understanding non-equilibrium self-organization, with implications for programmable active materials and synthetic or biological assemblies. While most prior studies have focused on dimers or small aggregates, the dynamics of larger clusters remain relatively unexplored. Here, we experimentally investigate chemically active, monodisperse \ce{SiO2}-Pt Janus colloid (JC) clusters as large as $n=9$ in a dynamic clustering regime, where clusters continuously form, dissolve, and merge as swimmer density increases. We show that clusters move in circular trajectories, and that both their translational and rotational dynamics can be predicted directly from the orientations of constituent JCs. Furthermore, we identify that their formation undergoes a mechanistic transition: while small clusters are mediated by chemical interactions, larger clusters are predominantly formed by steric effects. This transition arises from a mismatch of motilities between incoming JCs and clusters, combined with increased Pt-surface exposure. Our results extend prior dimer-focused studies to larger aggregates and establish a predictive description that bridges individual swimmer behavior with collective dynamics.

\end{abstract}

\maketitle

\section{Introduction}

``Artificial microswimmers'' refers to microscopic entities that convert chemical or other forms of energy into mechanical motion, and have emerged as a focal point of research due to their various characteristics and potentials. These synthetic entities are directly inspired from biological microorganisms, exhibiting greater motility and controllability compared to biological microswimmers \cite{Lauga2012}. Over the past years, researchers have developed artificial microswimmers of various sizes, shapes, and propulsion mechanisms \cite{Ebbens2016,Dwivedi2022}, driven by the goal of harnessing them for real-world applications. 

The propulsion of artificial microswimmers results from the asymmetry on the swimmer's surface, which creates and responds to gradients in the surrounding medium. The asymmetry enables the microswimmers to perform an autonomous motion, making them suitable for applications in various fields, such as targeted drug delivery \cite{Ma2015, Safdar2017}, microrobotics , environmental remediation \cite{Soler2014}, lab-on-a-chip technologies \cite{Garcia2013,Perez2014}, biomedical imaging and sensing \cite{Yang2012,Yi2016}, among others. Beyond practical uses, they also serve as model systems for studying not only the dynamics of biological microswimmers but also other naturally observed behaviors such as flocking \cite{Das2024}, pattern formation \cite{Liebchen2015,Das2024}, and predator-prey interactions \cite{Si2023}. 

One of the most common classes of artificial microswimmers is ``Self-propelled Colloids'', primarily consisting of Janus Colloids (JCs). Named after the two-faced Roman god \emph{Janus}, these colloids possess two distinct surface properties. When  JCs are dispersed in a suitable medium, and subjected to a driving force, the two hemispheres of a JC react asymmetrically with the surroundings, thus producing a self-generated gradient. These gradients drive the motion of the JCs, widely known as \emph{self-phoretic motion}. The generated gradient can be ionic (self-electrophoresis) \cite{Nourhani2015}, thermal (self-thermophoresis) \cite{Jiang2010,Avital2021} or chemical (self-diffusiophoresis), which is focus of the present study.

Self-diffusiophoretic JCs typically consist of an inert core, such as  \ce{SiO2} or polystyrene, half-coated with a catalytic metal like Pt or Au. When dispersed in an aqueous hydrogen peroxide (\ce{H2O2}) solution, the metallic side catalyses the decomposition of \ce{H2O2} into water (\ce{H2O}) and oxygen (\ce{O2}), creating a local concentration gradient that propels the particle either towards or away from the reacting side, depending on the solute-surface interactions \cite{Howse2007,Moran2017}. In the case of \ce{SiO2}-Pt JCs in \ce{H2O2}, the interactions between the solute (\ce{O2}) and the Pt surface are repulsive, causing the JC to move away from the Pt-side \emph{i.e.,} \ce{SiO2}-side forward. Other types of self-diffusiophoretic JCs also exist, such as Au-Pt JCs, where both hemispheres are reactive in \ce{H2O2} \cite{Theurkauff2012,Ginot2018}, and \ce{SiO2}-C JCs in a lower-critical mixture of water and 2,6-lutidine, where the carbon hemisphere is heated through a laser beam, causing local demixing of the solution, resulting in a concentration gradient of 2-6-lutidine around the JC \cite{Buttinoni2013,Gomez2016}. Numerous studies have been made to uncover the interactions in chemically active JCs \cite{Liebchen2019,Katuri2021,Liebchen2021,Singh2022,Raman2023,Sharan2023,Singh2024}

One of the most intriguing problems in active matter concerns the collective behavior of self-diffusiophoretic colloids, namely how their dynamics evolve as particle concentration increases. Over the past decade, extensive experimental and theoretical efforts have uncovered a rich variety of collective states in such systems \cite{Bechinger2016}. Unlike bacterial suspensions, which develop large-scale coherent motion, vortices, and turbulence at high densities \cite{Rabani2013}, self-diffusiophoretic colloids typically exhibit \emph{Dynamic Clustering}, wherein particles spontaneously form finite-sized clusters that continuously assemble, break apart, and reorganize \cite{Theurkauff2012,Palacci2013,Buttinoni2013,Ginot2018}. This behavior has been attributed to chemically mediated interactions \cite{Theurkauff2012} or to steric self-trapping arising from persistent propulsion \cite{Buttinoni2013}. Upon further increasing particle concentration, these systems may undergo phase separation into a dense clustered phase and a dilute gas-like phase, reminiscent of motility-induced phase separation (MIPS) \cite{Cates2015,Gonnella2015MIPS}. At the same time, studies have shown that chemical interactions can suppress or modify MIPS, giving rise to arrested or oscillatory states instead \cite{Liebchen2015,Zhao2023}. Both experiments and simulations further report that average cluster size increases with particle activity, often characterized by the P\'eclet number $Pe$ \cite{Theurkauff2012,Buttinoni2013,Kalil2021}. Complementary numerical studies have mapped out phase diagrams based on translational and rotational phoretic interactions, identifying regimes such as active gases, dynamic clustering, chemotactic collapse, and phase separation \cite{Saha2014,Pohl2014,Pohl2015,Fadda2023}.

Despite this progress, most existing studies focus on how clusters form and on their steady-state size statistics, while the dynamics of the clusters themselves, and their connection to the orientations of the constituent JCs, remain comparatively unexplored. Some insight has been gained for induced-charge electrophoretic Janus colloids, where small clusters (up to four particles) were shown to behave approximately as rigid bodies, with their linear and angular velocities determined by the orientations of individual particles \cite{Boymelgreen2014,Biswas2025}. For chemically active Janus colloids, however, such configuration-dependent cluster dynamics have been investigated only at the level of dimers. In a seminal study, \citet{Ebbens2010} provided a qualitative description of dimer motion in Polystyrene–Pt Janus colloids, suggesting that parallel alignment leads to persistent motion while misalignment induces cyclic trajectories. Beyond dimers, experimental information on internal orientations remains limited. While \citet{Buttinoni2013} demonstrated that Janus colloids at the perimeter of sterically formed clusters predominantly point inward, their focus was on cluster formation rather than cluster motility. Subsequently, \citet{Ginot2018} showed that a minimal model assuming random orientations could nevertheless reproduce average cluster speeds, leaving open the question of how internal configuration and dynamics are related at the level of individual clusters.

At the same time, the chemically polar nature of Janus colloids is known to strongly influence pairwise interactions, collision geometries, and reorientation dynamics \cite{Katuri2021,Sharan2023,Singh2024}. These microscopic interaction rules directly affect how particles attach to clusters, how orientations evolve within clusters, and ultimately how clusters move. Yet, the links between interparticle interactions, cluster formation and growth mechanisms, internal orientational order, and emergent cluster dynamics are often treated separately. In this study, we address this gap by using chemically active \ce{SiO2}–Pt Janus colloids, propelled in a 0.5\% aqueous \ce{H2O2} solution, as a model system to systematically investigate cluster dynamics across a wide range of cluster sizes and particle concentrations. By directly comparing the motion of isolated particles and clusters, and by explicitly resolving the orientations of individual Janus colloids within clusters, we extend existing frameworks beyond dimers to elucidate how internal configuration governs cluster motion and how the dominant clustering mechanism evolves with cluster size.

\section{Experimental Methodology}

\subsection{Janus Particle Synthesis}
A stock solution of bare \ce{SiO2} microspheres (Sigma Aldrich, diameter $2a=5 \mu{m}$, 5 wt.\% solids) is diluted to one-fourth of its original concentration using ethanol. Subsequently, 80 $\mu$l of the diluted suspension was deposited onto an \ce{O2}-plasma-treated glass slide (22 mm $\times$ 22 mm, PDC-32-G2, Harrick Plasma) and spin-coated (EZ1, Apex) at 2000 rpm for 90 seconds, following an acceleration period of 210 seconds. This process produced a uniform monolayer of \ce{SiO2} microspheres on the glass slide, which was confirmed using optical microscopy. 

The monolayer was then coated with a thin layer ($\sim$ 15 nm) of Pt through a DC Magnetron Sputter coater (BT150, Hind High Vacuum). The self-shadowing effect results in the coating of only the top half of the microspheres, thus rendering them as Janus spheres. The coated sample is then sonicated in ultrapure water for 5 minutes, resulting in a JC suspension. The JCs are then washed with water by centrifuging twice at 6500 rpm for 5 minutes, and subsequently they were re-dispersed in ultrapure water. The resultant stock solution is then appropriately diluted to achieve the desired particle concentration for the experiments. To make the particles chemically active, an appropriate volume of 30 \% (w/v) solution of Hydrogen peroxide (\ce{H2O2}, Qualigens) is added to the dilute suspension to achieve a \ce{H2O2} concentration of 0.5\% (w/w).

\subsection{Sample preparation and Imaging}
To image the JCs' motion, a poly(dimethylsiloxane) (PDMS, Sylgard 184, Dow Chemical) spacer of dimensions 25 mm $\times$ 25 mm $\times$ 7 mm with a cavity of diameter 10 mm in the middle was created using suitable molds. A crosslinker amount of around 5 wt.\% as opposed to the recommended 10 wt.\% was used to ensure proper adhesion of the PDMS spacer to the glass substrate. Larger than usual cavity size was used to mitigate the effects of any advection caused due to \ce{O2} bubbling. Glass slides are sonicated in Isopropyl alcohol for 20 min., followed by an \ce{O2}-plasma treatment for 5 min. The particle suspension is then added to the cell cavity and then sealed with a glass coverslip from the top. Due to density effects, the JCs settle to the bottom of the well and perform a quasi 2-D motion, in their characteristic `half-moon' orientation\cite{Das2015} upon addition of adequate \ce{H2O2} to the aqueous phase. Once all the particles are settled at the bottom of the well, imaging was done using an inverted microscope (IX73, Olympus) with a 20$\times$ objective in bright-field mode. The microscope was coupled with a CMOS Camera (Oryx 10GigE, Teledyne FLIR) of resolution 2048 $\times$ 2048 pixel$^2$. Videos were recorded for 120 s at 2 Hz.

\subsection{Particle Tracking and Data Analysis}
From the recorded videos, the center-of-mass positions [$x(t),y(t)$] of the particles and clusters are extracted using the Trackmate \cite{Tinevez2017} plugin of the image processing tool Fiji \cite{Schindelin2012}. Mask detector was used to detect the centers of larger (Number of particles $n \geq 2$) clusters. For detecting the centres of single particles and their Pt-hemispheres, the Laplacian-of-Gaussian (LoG) detector was used. Subsequently, the trajectories of the particles and clusters are obtained using the Linear Assignment Problem (LAP) tracker algorithm\cite{Jaqaman2008}. All the detection and tracking algorithms are inbuilt into the plugin. Further analysis using the ($t,x,y$) data was done using custom-written codes in MATLAB and Python.

\section{Results and Discussion}
\subsection{Control Experiments: Motion of isolated 5 $\mu$m active JCs}
\begin{figure*}[t]
    \centering
    \includegraphics[width=\textwidth]{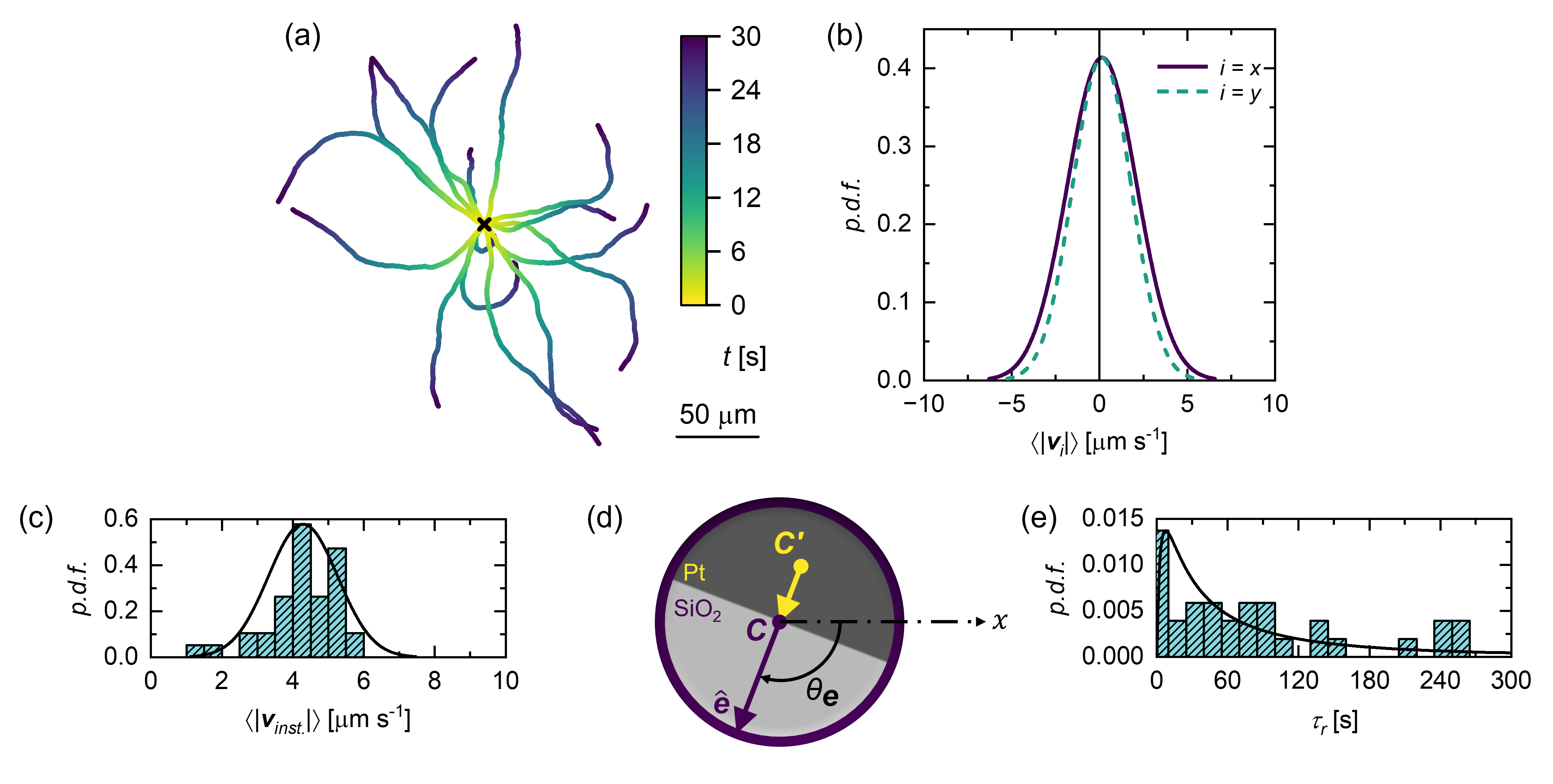}
    \caption{\textbf{Dynamics of isolated active \ce{SiO2}-Pt JCs:} (a) Representative trajectories of a few isolated active \ce{SiO2}-Pt JCs in 0.5\% \ce{H2O2} solution. The cross mark refers to the JCs' initial position. (b) Fitted gaussian curves of the projected velocity components $\langle|\textbf{v}_x|\rangle$ and $\langle|\textbf{v}_y|\rangle$ of 40 isolated active JCs. (c) Probability distribution of the average instantaneous speed $\langle|\textbf{v}_{inst.}|\rangle$ of the active JCs with a suitable gaussian fit (solid line). (d) Schematic of an active JC depicting the orientation unit vector $\hat{\boldsymbol{e}}$, orientation angle $\theta_\textbf{e}$, centre of JC ($C$), and the centre of the Pt hemisphere ($C'$). (e) Probability distribution of the Brownian Rotational Timescale $\tau_{r}$ of the active JCs with a suitable lognormal fit (solid line).}
    \label{fig:Control}
\end{figure*}

To establish a baseline understanding, we examined the motion of 40 active 5 $\mu$m \ce{SiO2}-Pt JCs in 0.5 wt.\% aqueous \ce{H2O2} solution in highly dilute environments, \emph{i.e.,} area fraction of JCs $\sim 0$. Figure \ref{fig:Control}(a) shows representative two-dimensional ($x$,$y$) trajectories of a few active JCs recorded over 30 s. As seen in the figure, the particles move persistently in various directions, indicating the absence of any advection or fluid leakage within the optical cell. Further confirmation is provided in figure \ref{fig:Control}(b), showing the probability distributions of the magnitudes of the velocity components $|\textbf{v}_i|$ ($i$=$x$,$y$) of the JCs. Both $|\textbf{v}_x|$ and $|\textbf{v}_y|$ follow normal distributions centered around zero, corroborating the lack of directional bias in the system. The active JCs move with a mean speed of $\langle|\textbf{v}_{inst.}|\rangle$ $=$ 4.28 $\pm$ 0.97 $\mu$m s$^{-1}$. Figure \ref{fig:Control}(c) shows the distribution of $\langle|\textbf{v}_{inst.}|\rangle$ along with a Gaussian fit. 

The instantaneous orientation of JC is represented by the unit vector $\hat{\boldsymbol{e}}$ drawn normal to the Janus equatorial plane, as illustrated in figure \ref{fig:Control}(d). The angle $\theta_\textbf{e}$ is defined as the angle between $\hat{\boldsymbol{e}}$ and the spatial $x$-axis, in the lab frame. Experimentally, $\theta_\textbf{e}$ is calculated from the slope of the line connecting the center of the Pt-hemisphere ($C'$) with the center of the JC ($C$) \cite{Zhou2023}. Both these points are extracted from ImageJ. From the obtained orientation unit vector angles, we compute the Orientation Autocorrelation Function (OACF) $C_{\theta\theta}({\Delta}t)=\langle\cos{[\theta_{\boldsymbol{e}}{(t)}-\theta_{\boldsymbol{e}}{(t+{\Delta}t)}]}\rangle \propto \cos{(\omega\Delta{t})}.\exp{-\left(\frac{\Delta{t}}{\tau_{r}}\right)}$, where $\tau_{r}$ is the Brownian Rotational Timescale and $\omega$ represents the angular speed of the JC, reflecting its cyclic behavior (See Supplementary Material, Figure S1(a)). The probability distribution of $\tau_{r}$ of the isolated JCs, shown in figure \ref{fig:Control}(e), is well described by a lognormal fit, with an average value $\tau_{r}$ of 65.75 $\pm$ 46.37 s. Furthermore, we calculate the two-dimensional Mean Square Displacement (MSD) $\langle{\Delta}r^2\rangle = \langle{\Delta}x^2\rangle + \langle{\Delta}y^2\rangle$ of the active JCs (See Supplementary Material, Figure S1(b)). At shorter timescales ($\Delta{t} \ll {\tau}_r$), the MSD curves follow a power law exponent of 2, indicating the ballistic nature of the JCs, and the curves agree well with the form $\langle{\Delta}r^2\rangle$ (${\Delta}t$) = $v^2{\Delta}t^2+4D{\Delta}t$ for $\Delta{t}\ll\tau_{r}$, where $v$ is the phoretic speed and $D$ is the Brownian diffusivity (See Supplementary Material, Figure S1(c)). 

\subsection{Effect of Increasing particle population}
\begin{figure*}[t]
    \centering
    \includegraphics[width=\textwidth]{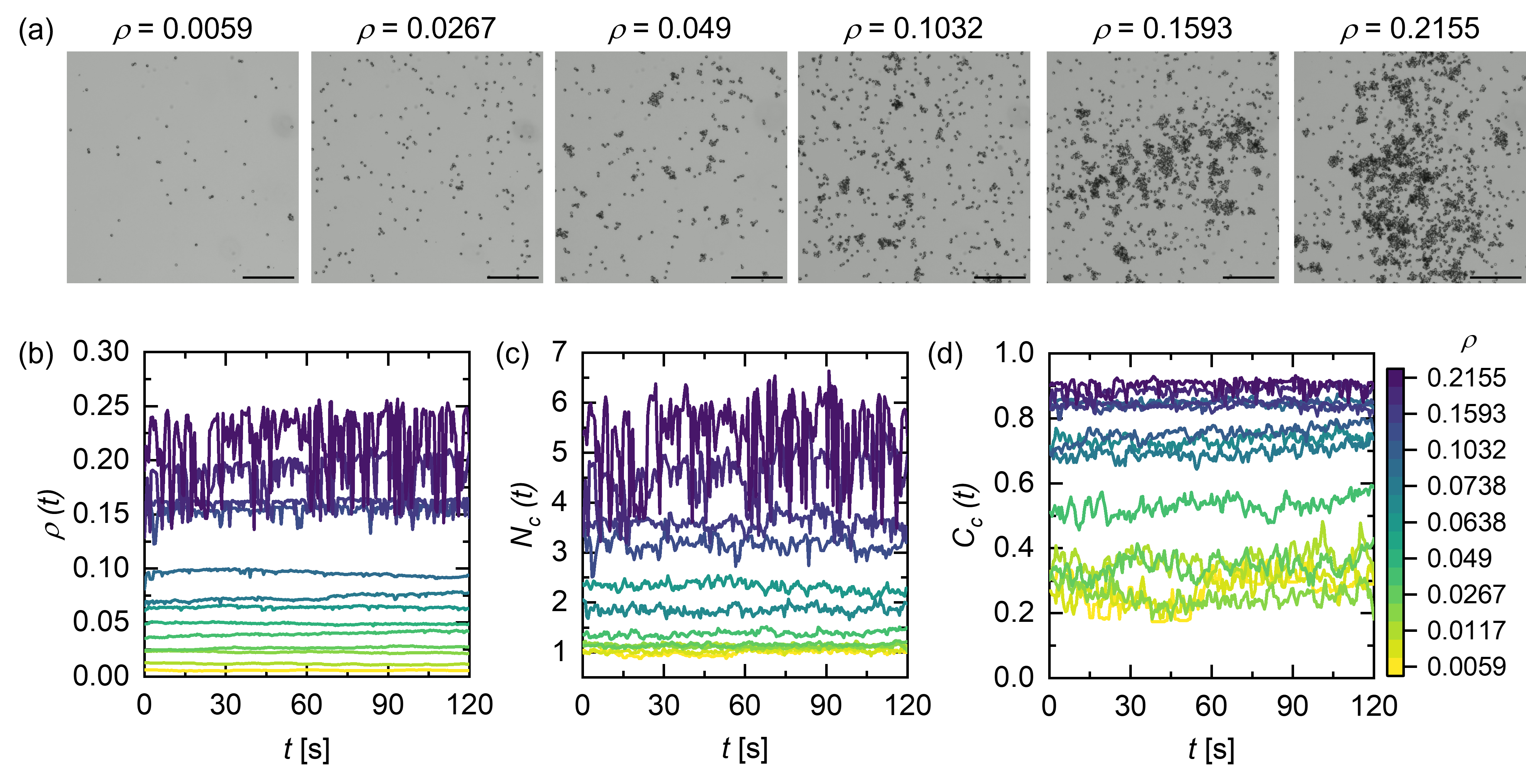}
    \caption{(a) Optical micrographs showing the system of active JCs at various particle area fraction $\rho$. Scale bars indicate a length of 100 $\mu$m. (b-d) Temporal variation of Area fraction $\rho$, Average cluster size $N_c$, and Clustering coefficient $C_c$, respectively, across systems of various particle concentrations. For a clearer visualization, some of the curves have been omitted.}
    \label{fig:Invariance}
\end{figure*}

Building on the observations of active JCs in dilute environments, we systematically increase the particle population, \emph{i.e.,} the area fraction of JCs. As shown in figure \ref{fig:Invariance}(a), as the area fraction increases, particles get closer to each other and form clusters. Different clusters form, merge, reform, and dissolve simultaneously (see Supplementary Material, video S1), resulting in the well-known ``dynamic clustering'' state. 

Since our observation window (roughly 450 $\mu$m $\times$ 450 $\mu$m) is much smaller than the dimensions of the optical well, it is necessary to show that the statistics obtained within this window are sufficient to characterize the entire system. We do so by checking if the particle area fraction within the observation window remains constant over time, \emph{i.e.,} no net influx or ouflux of particles occurs within the observation window. We observe this over a window of 120 s. At very low particle area fractions, the Brownian rotational timescale $\tau_{r}$ of isolated particles is ~60 s, allowing the JCs to reorient within the observation window. At higher fractions, one would expect significantly higher collision frequencies with neighboring JCs, leading to their decorrelation of motion at much shorter timescales. Hence, an observation window of 120 s is sufficient to capture statistically meaningful behavior across across area fractions. 

At any time $t$, we define the area fraction $\rho(t)$ as $\rho(t)=\frac{N(t) \times A_p}{A_W}$, where $N(t)$ is the number of particles in the observation window at the given time instant, and ${A_p}$ and $A_{W}$ denote the projected area of a single particle and the area of the observation window, respectively. Figure \ref{fig:Invariance}(b) illustrates the variation of $\rho(t)$ for systems with varying particle concentrations. Notably, in all cases, the area fraction within the observation window remains time-invariant, displaying only minor fluctuations. Consequently, throughout this study, $\rho$ represents the time-averaged area fraction. Herein, we report the results up to $\rho=0.2155$, as particle concentrations greater than that led to vigorous bubbling, effectively disrupting the visualization of the JCs. 

To further characterize the clustering dynamics across area fractions, we examine the temporal behavior of the average cluster size $N_c$ and the clustering coefficient $C_c$. Figure \ref{fig:Invariance}(c) depicts the temporal variation of the average cluster size $N_c$, for systems of different $\rho$. Similar to $\rho$, $N_c$ remains time-invariant while increasing with particle density. This invariance implies that the processes of splitting, merging and exchanging particles within clusters occur concurrently, effectively canceling each others' effects. As a result, the global average remains stable over time despite the short-term fluctuations, \emph{i.e.}, a quasi-steady-state behavior. A similar behavior was observed for the clustering coefficient $C_c$, which is defined as the fraction of particles in the system that belong to a cluster. At any timestamp $t$, $C_c$ is calculated as:
\begin{equation}
    C_c=\frac{\sum_{i=2}^{\infty} iN_i}{\sum_{i=1}^{\infty} iN_i},
\end{equation} 
where $N_i$ denotes the number of clusters of size $i$ in the observation window. $C_c \rightarrow 0$ means that all the JCs in the observation window are isolated \emph{i.e.,} gaseous phase, while $C_c \rightarrow 1$ indicates that the JCs are never isolated \emph{i.e.,} they are part of one of the clusters. As shown in figure \ref{fig:Invariance}(d), $C_c$ exhibits only minor fluctuations while remaining nearly constant over time. This consistency further supports the dynamic clustering behavior of the active JCs and not static aggregation.

\begin{figure*}
    \centering
    \includegraphics[width=0.8\textwidth]{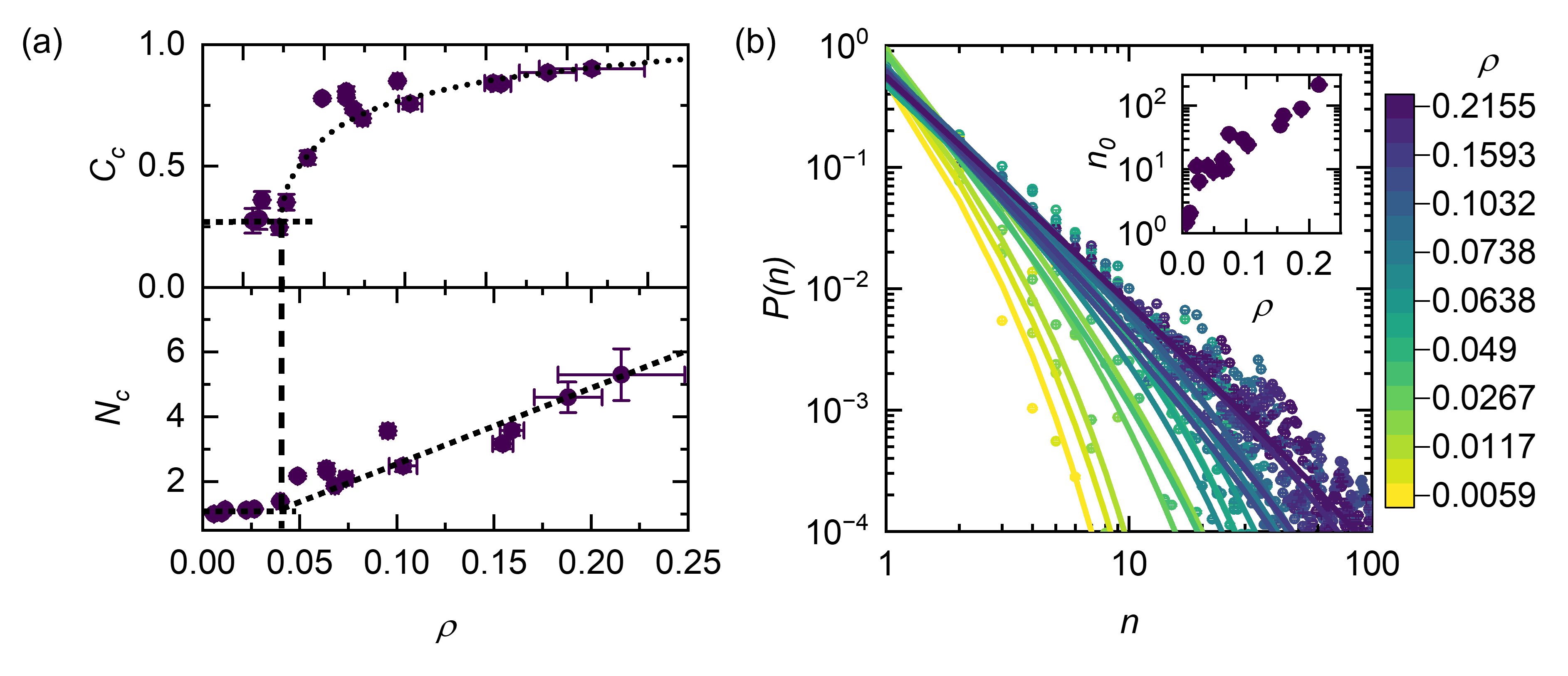}
    \caption{(a) Area fraction dependence of the (time-averaged) Clustering coefficient $C_c$ and the Average cluster size $N_c$. Dotted lines are drawn as eye guides for the reader. (b) Cluster Sizewise distribution of (time-averaged) Probability $P(n)$ and for systems of various $\rho$, based on averaging over a period of $\sim$ 120 s. Solid lines indicate the corresponding function fits. Inset shows the variation of the characteristic cutoff cluster size $n_0$ with area fraction. Error values are smaller than the marker size and hence not visible.}
    \label{fig:Clustering}
\end{figure*}

Figure \ref{fig:Clustering}(a) illustrates the variation of the clustering coefficient $C_c$ and average cluster size $N_c$ with particle fraction $\rho$. Our observations indicate that for $\rho \lesssim 0.025$, $C_c$ remains relatively constant at $\sim$ 0.1, while $N_c$ stays close to 1. This behavior is expected, as clustering events are rare at such low particle concentrations. The corresponding low values of $C_c \sim 0.1$ reflects the presence of small aggregates, primarily two-particle clusters (dimers) and occasional trimers. Beyond $\rho \sim 0.025$, both $C_c$ and $N_c$ increase monotonically, indicating the onset of clustering. While $N_c$ shows a linear growth over the range of concentration studied, $C_c$ increases non-linearly and tends to plateau at higher $\rho$. Similar growth and plateauing has been predicted for an ensemble of active inertial particles interacting via a Lennard-Jones potential \cite{Cantisan2023}.

This trend is consistent with the previous findings by Buttinoni \etal, who investigated the clustering dynamics of \ce{SiO2}-Graphite ACs and showed qualitatively that increasing $\rho$ beyond a threshold (around $\rho \sim 0.25$) leads to a phase separation transition consisting of a few clusters and some free JCs (\emph{i.e.,} $C_c \sim 1$). They also showed through simulations that when $\rho$ is increased further, most particles coalesce into a single giant cluster (\emph{i.e.,} very high $N_c$)\cite{Buttinoni2013}. While the fraction of particles in our experiments is lower compared to theirs, qualitatively we see a similar observation of growth of larger clusters, as seen from the cluster size distribution $P(n)$, at different values of $\rho$ shown in figure \ref{fig:Clustering}(b). At any time point, $P(n)$ is computed by dividing the frequency (count) of clusters of size $n$ by the total number of clusters. 
\begin{equation}
    P(n)=\frac{N_n}{\sum_{i=1}^{\infty} N_i}
\end{equation}
where $N_n$ denotes the number of clusters of size $n$ in the observation window. 
We observe that the (time-averaged) $P(n)$ broadens with increasing $\rho$. At low particle concentrations, larger clusters are rare, but their likelihood increases at higher $\rho$. In addition, the overall shape of $P(n)$  curves closely resembles those obtained in previous experimental and computational works \cite{Buttinoni2013,Pohl2015}, and can be represented by a truncated power-law form:
\begin{equation}
    P(n) \propto n^{-\beta} \exp\left(-\frac{n}{n_0}\right)
    \label{Prob}
\end{equation}
with a power-law exponent $\beta=2.05~\pm~0.25$ and $n_0$, a characteristic cutoff cluster size, increasing with $\rho$, as shown in the inset of figure \ref{fig:Clustering}(b). The truncated power-law form of $P(n)$ suggests that clusters in our system do not grow indefinitely and coarsen beyond a characteristic size $n_0$, indicating the presence of an effective repulsive mechanism that limits aggregation. The measured exponent, $\beta \sim 2$, is in close agreement with the predictions of \citet{Pohl2015} for active systems with repulsive (rotational) phoretic interactions, who report $\beta = 2.1 \pm 0.1$ at low particle densities. Furthermore, our $P(n)$ curves exhibit a single-step decay, consistent with the ``Dynamic Clustering 1'' state identified in their study. In contrast, upon increasing interparticle attraction, \citeauthor{Pohl2015} observe a second dynamic clustering regime characterized by a larger mean cluster size and a clear inflection in $P(n)$, which can be described by a superposition of two truncated power-law contributions. The absence of such inflection points in our $P(n)$ curves, together with the close agreement of the exponent and decay form, indicates that our system remains within a repulsion-dominated dynamic clustering regime analogous to their Dynamic Clustering 1 state, with no evidence of a crossover to a more strongly attractive clustering phase.

However, unlike previously investigated phoretic systems in which both hemispheres are catalytically active and clustering is driven primarily by long-range chemical fields \cite{Theurkauff2012,Ginot2018}, cluster formation in our system can additionally proceed via a steric pathway owing to the inertness of the \ce{SiO2} hemisphere. In particular, contacts between the \ce{SiO2} hemispheres of neighboring JCs can give rise to transient steric trapping, analogous to the self-trapping mechanism described by Buttinoni \etal \cite{Buttinoni2013}, where persistent propulsion and hindered reorientation lead to particle accumulation even in the absence of explicit attractive interactions. While such steric effects can facilitate local aggregation and cluster nucleation, they do not, by themselves, imply long-term coarsening or permanent arrest.

Within the theoretical framework proposed by \citet{Saha2014}, our system is best described as operating in a reaction-limited regime (high reactant concentration), with particles exhibiting a positive chemotactic response to the reactant ($A>0$) and an effective phoretic repulsion from the product ($B<0$). This combination is predicted to yield finite-sized clusters and inward-pointing aster-like structures. Similar behavior was also reported by \citet{Liebchen2015}, where repulsive chemical interactions prevent indefinite coarsening while still allowing dynamic clustering. While our observations qualitatively align with these predictions in that clustering occurs without attractive interactions and remains dynamically fluctuating, the absence of a secondary decay or inflection in our $P(n)$ curves suggests that the system does not enter a distinct, strongly arrested clustering regime. Instead, clusters span a broad range of sizes without a sharply defined saturation scale. To further elucidate the underlying mechanism, we therefore next analyze the dynamics of clusters and examine how cluster motility scales with cluster size.

\subsection{Cluster Dynamics}

After establishing the structural features of clustering using metrics like $C_c$, $N_c$, and $P(n)$, and proposing a clustering mechanism, we now investigate the dynamical aspects of cluster formation. Our specific goal is to comprehend how cluster motility changes as a function of cluster size and what this indicates about the clustering mechanism.

\begin{figure*}[t]
    \centering
    \includegraphics[width=\textwidth]{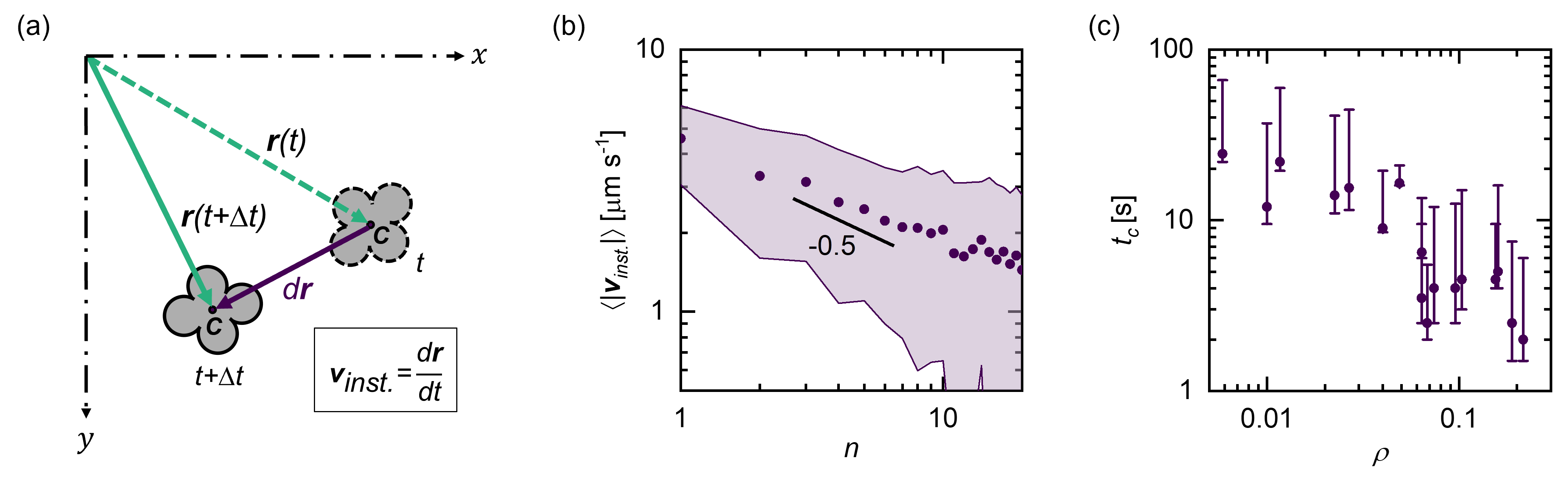}
    \caption{(a) Schematic illustrating the estimation of the instantaneous velocity vector $\textbf{v}_{inst.}$ for an arbitrary cluster at a given instant. (b) Variation of mean $\langle|\textbf{v}_{inst.}|\rangle$ with cluster size $n$ in logarithmic scale with combined data across all $\rho$. (c) Variation of the median cluster size retention time $t_c$ with area fraction $\rho$, combining clusters of all sizes. Error bars denote the interquartile range.}
    \label{fig:Clusterdynamics1}
\end{figure*}

We begin by examining the average instantaneous speed, $\langle|\textbf{v}_{inst.}|\rangle$, of isolated Janus colloids (JCs) and clusters across different area fractions $\rho$. Similar to a single JC, the instantaneous velocity vector $\textbf{v}_{inst.}$ for a cluster is defined as $\textbf{v}_{inst.}= \dv{\textbf{r}}{t} =\frac{\textbf{r}(t+{\Delta}t)-\textbf{r}(t)}{{\Delta}t}$, where $\textbf{r}(t)$ is center-of-mass position vector of the JC (cluster) and ${\Delta}t$ ($=$ 0.5 s in our case) is the time step, as illustrated in figure \ref{fig:Clusterdynamics1}(a). We did not observe any consistent dependence of $\langle|\textbf{v}_{inst.}|\rangle$ on $\rho$ for any cluster size (see Supplementary Material, Figure S2(a)). While a previous study reported a reduction in single-particle speeds at low $\rho$, attributed to an increased rate of pairwise interactions \cite{Singh2024}, our results indicate that this effect is screened beyond a certain density. Therefore, we categorize clusters based solely on their size, independent of system density. Figure \ref{fig:Clusterdynamics1}(b) presents the variation of $\langle|\textbf{v}_{inst.}|\rangle$ with cluster size for $n \leq 20$, which exhibits a clear scaling behavior of $\langle|\textbf{v}_{inst.}|\rangle \propto n^{-0.5}$.

The observed scaling behavior is consistent with established models of phoretic motion, where particle speed decreases with increasing effective radius as $v \propto a^{-1}$.\cite{Ebbens2012} The projected area of a cluster can be expressed as $A_p = \pi{r^2} = n\times\pi{a^2}$, where $a$ is the radius of a single JC and $r$ is the projected area-equivalent radius. This yields $r \propto n^{0.5}$, consistent with the observed scaling trend in $\langle|\textbf{v}_{inst.}|\rangle$. 

Figure \ref{fig:Clusterdynamics1}(c) shows the variation of the median cluster size retention time, $t_c$, with area fraction $\rho$. The size retention time is defined as the duration over which a cluster retains a constant number of constituent particles, \emph{i.e.,} no particle joins or leaves the cluster. We observe a clear decreasing trend of $t_c$ with increasing particle fraction $\rho$, and unlike the average instantaneous speed $\langle|\textbf{v}_{inst.}|\rangle$, $t_c$ exhibits no discernible dependence on cluster size (see Supplementary Material, Figure S2(b)). These trends in $t_c$ can be attributed to the heightened frequency of particle interactions at higher $\rho$, which promotes both cluster merging and fragmentation. Thus, the increasingly dynamic environment at higher densities renders the lifespan of clusters shorter, despite their potential to attain larger sizes.

We now shift our focus towards the dynamics of isolated clusters to better understand their motility characteristics. For this analysis, we carefully selected $\sim$ 30 clusters for each $1 < n < 10$, selected based on a minimum size retention time of $t_c \geq 15$ s at the lowest accessible area fraction. These constraints help ensure that the clusters are both stable and dynamically isolated during observation.
Figure \ref{fig:Orientation}(a-d) shows the optical micrographs of a single JC and three representative clusters of different sizes at time $t = 0$, with their full trajectories overlaid (see Supplementary Material, videos S2-S5). Notably, in contrast to an isolated JC, the clusters display circular or helical trajectories, indicative of persistent rotational dynamics. Previously, circular motion has been observed for chemically active, isolated JCs with imperfect metallic coating \cite{Wang2017} and JCs in viscoelastic media \cite{Narinder2018}. \citet{Ebbens2010} observed cyclic trajectories in dimers of Pt-Polystyrene active JCs in aqueous \ce{H2O2} media, consistent with our observations. However, their work is limited to dimers and did not provide a direct visualization of particle orientations during the cyclic motion, relying instead on schematic illustrations to describe the trajectories. In contrast, our work not only visualizes the full orientational dynamics but also extends to assemblies beyond dimers.

\begin{figure*}[t]
    \centering
    \includegraphics[width=\textwidth]{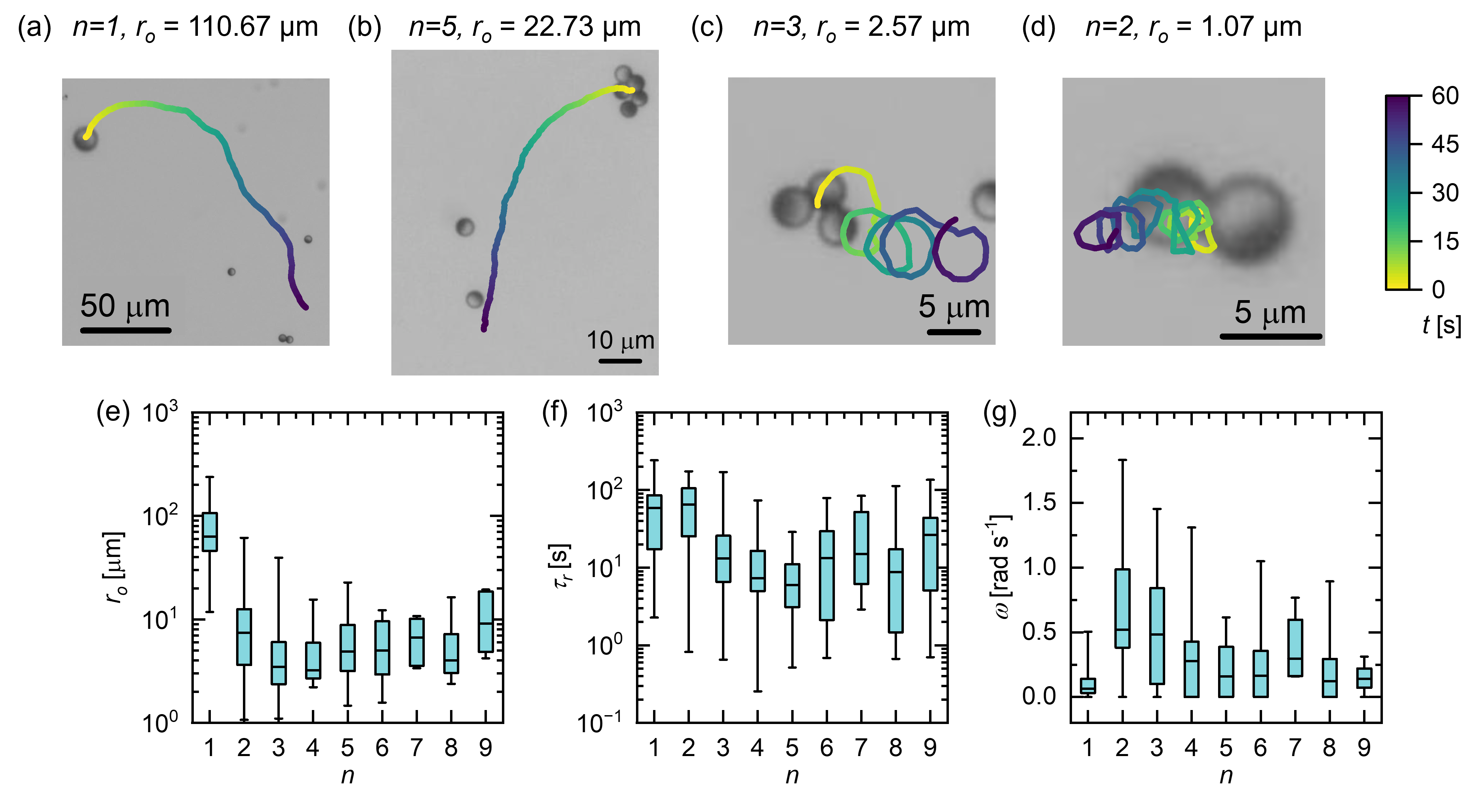}
    \caption{Trajectories (duration $\sim$ 60 s) of (a) a single JC and (b-d) a few representative clusters with different sizes and orbit radii, superimposed on their optical micrographs at time $t=0$. The JC shown in figure (a) is enlarged for better visualization and hence not to scale. (e-g) Variation of the orbit radius $r_o$, Brownian Rotational timescale $\tau_r$, and Angular speed $\omega$, respectively, for a group of nearly 30 clusters at each $n$. Bar denotes the median, box denotes the inter-quartile range, and whiskers denote the minimum and maximum values.}
    \label{fig:Clusterdynamics2}
\end{figure*}

To characterize the clusters' cyclic motion, we estimate the orbit radius, $r_o$, of each cluster by fitting a single cycle of its trajectory to a circular arc. For trajectories that do not exhibit cyclic behavior, we approximate the entire trajectory using a circle, interpreting linear trajectories as a limiting case of cyclic trajectories with very large $r_o$ values.

Figure \ref{fig:Clusterdynamics2}(e) presents the variation of $r_o$ with cluster size $n$. While isolated JCs ($n=1$) typically exhibit large orbit radii due to their persistent translational motion, we find that the median $r_o$ for clusters is considerably lower. Interestingly, beyond this initial drop, $r_o$ does not exhibit significant dependence on cluster size. This absence of a clear trend is likely a consequence of the structural diversity among clusters of a given size including variations in particle arrangement, alignment, and internal torque that can all lead to wide variations in trajectory shape. These configurational effects will be examined in detail in the next subsection. 

We further compute the MSD and OACF of these clusters, and fit them to their respective analytical forms. For OACF, we use the angle of the instantaneous velocity vector $\theta_{\boldsymbol{v}}$ to determine orientation persistence, and the OACF is defined as 
\begin{equation}
    C_{\theta\theta}=\langle\cos{[\theta_{\boldsymbol{v}}{(t)}-\theta_{\boldsymbol{v}}{(t+{\Delta}t)}]}\rangle
\end{equation}
Both the MSD and OACF exhibit oscillatory behavior (see Supplementary Material, figure S3), a hallmark for circular or helical trajectories \cite{Ebbens2010}. From MSD and OACF, we extract the persistence timescale $\tau_p$ and the angular speed $\omega$ of the clusters. 

Figure \ref{fig:Clusterdynamics2}(f) shows how the persistence timescale $\tau_p$ with cluster size $n$. We observe a weakly decreasing trend, indicating that larger clusters reorient more rapidly and thus exhibit reduced persistence in their direction of motion, consistent with the cyclic motion of the clusters. The variation of the angular speed $\omega$ of the clusters is presented in figure \ref{fig:Clusterdynamics2}(g). As expected, single JCs show low angular speed due to their highly directed, ballistic motion. In contrast, 2- and 3-particle clusters display a pronounced peak in $\omega$, reflecting their strong cyclic dynamics. As the cluster size increases further, $\omega$ gradually decreases.

\subsection{Effect of Cluster configuration}

\begin{figure*}[t]
    \centering
    \includegraphics[width=\linewidth]{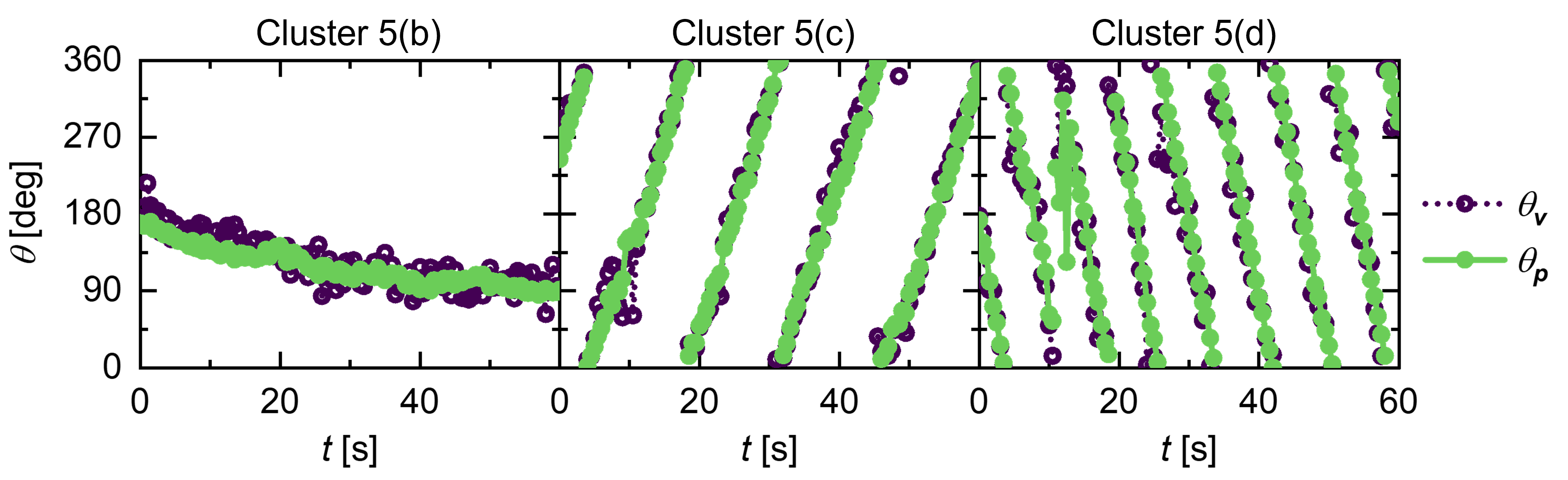}
    \caption{Temporal variation of the angles of the instantaneous velocity vector $\theta_{\textbf{v}}$ and the polarization vector $\theta_{\textbf{p}}$ for the trajectories shown in figure \ref{fig:Clusterdynamics2} (b-d).}
    \label{fig:Orientation}
\end{figure*}

In their earlier study, \citet{Ebbens2010} qualitatively demonstrated that the persistence of dimers correlate with the alignment of the participating JCs. Specifically, dimers with similarly oriented JCs tended to be more persistent and dimers with oppositely aligned JCs exhibited cyclic motion. This suggests that the configuration of active JCs within a cluster strongly influences its motion characteristics and thus could be the reason for most clusters exhibiting cyclic motion with varying orbit radii $r_o$in our experiments. To quantify the influence of participating JCs' configurations on the overall cluster dynamics, we define a polarization vector $\boldsymbol{p}$ for an $n$-particle cluster as:
\begin{equation}
    \boldsymbol{p}=\frac{1}{n}\sum_{i=1}^{n} \hat{\boldsymbol{e}}_i,
\end{equation}
where $\hat{\boldsymbol{e}}_i$ denotes the orientation unit vector of the $i$-th JC, as illustrated earlier in figure \ref{fig:Control}(d). We compare the polarization vector $\boldsymbol{p}$ with the instantaneous velocity vector $\boldsymbol{v}_\text{inst}$. Specifically, we track their angles relative to a fixed $x$-axis, denoted $\theta_{\boldsymbol{p}}$ and $\theta_{\boldsymbol{v}}$, respectively. The comparison, shown in figure \ref{fig:Orientation}, demonstrates a strong correlation between the two. This close alignment implies that the resultant motion of a cluster can be inferred from the average orientation of its constituent JCs. Just as in the case of an isolated JC, the net movement of a cluster is closely correlated with its net orientation. While a linear combination of the orientations of the constituent JCs is simple and straightforward, it still serves as an effective descriptor of a cluster's movement.

\begin{figure*}[t]
    \centering
    \includegraphics[width=\textwidth]{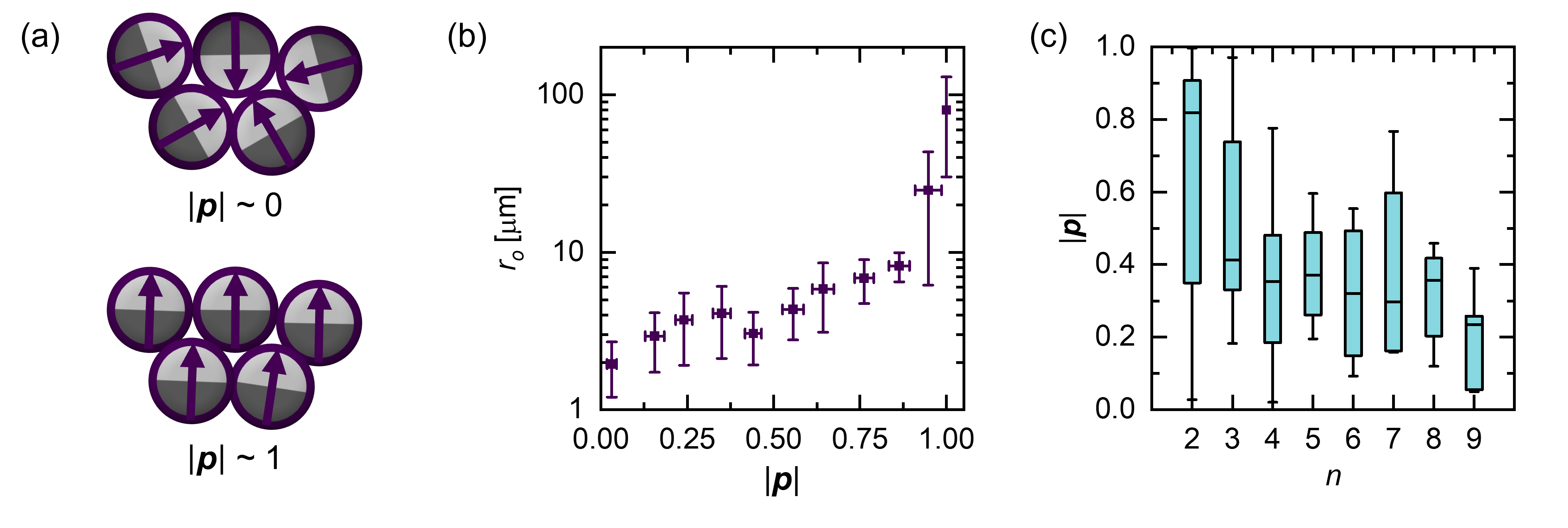}
    \caption{(a) Schematics illustrating a disordered ($|\boldsymbol{p}|=0$) and an ordered ($|\boldsymbol{p}|=1$) cluster (b) Variation of the orbit radius $r_o$ as a function of polarization vector $|\boldsymbol{p}|$ across single JCs and clusters upto size $n=9$. The data point at $|\boldsymbol{p}|=1$ correspond to isolated JCs. (c) Distribution of $|\boldsymbol{p}|$ for clusters of different sizes. Bar denotes the median, box denotes the inter-quartile range, and whiskers denote the minimum and maximum values.}
    \label{fig:Correlation}
\end{figure*}

The degree of orientational order can be captured by the magnitude of the polarization vector $|\boldsymbol{p}|$. By definition, $|\boldsymbol{p}|$ ranges from 0 to 1, with $|\boldsymbol{p}|=1$ corresponding to perfect alignment of all JCs, and $|\boldsymbol{p}|=0$ indicating complete orientational disorder, as depicted in figure \ref{fig:Correlation}(a). Through our experiments, we extend this understanding to larger clusters. Figure \ref{fig:Correlation}(b) illustrates the variation of $r_o$ with $|\boldsymbol{p}|$ for single JCs clusters upto size $n=9$. We observe that $r_o$ initially increases slowly with $|\boldsymbol{p}|$ ($\leq 0.8$) and sharply as $|\boldsymbol{p}|$ approaches $1$. This trend clearly suggest that higher orientational alignment (\emph{i.e.,} higher $|\boldsymbol{p}|$) of the participating JCs in a cluster leads to more persistent trajectories (larger $r_o$), and that the circular motion is inevitable in clusters consisting of misaligned JCs, as shown by the majority of the clusters.

The above correlations show that $|\boldsymbol{p}|$ can be used to describe both the direction of motion and the persistence of a cluster. To understand how the orientational order evolves with cluster size, we examine the distribution of $|\boldsymbol{p}|$ across different $n$, shown in figure \ref{fig:Correlation}(c). We find that dimers span the entire range of $|\boldsymbol{p}|$, whereas larger clusters show narrower distributions centered at smaller values, indicating progressive loss of alignment.

\begin{figure*}[t]
    \centering
    \includegraphics[width=\textwidth]{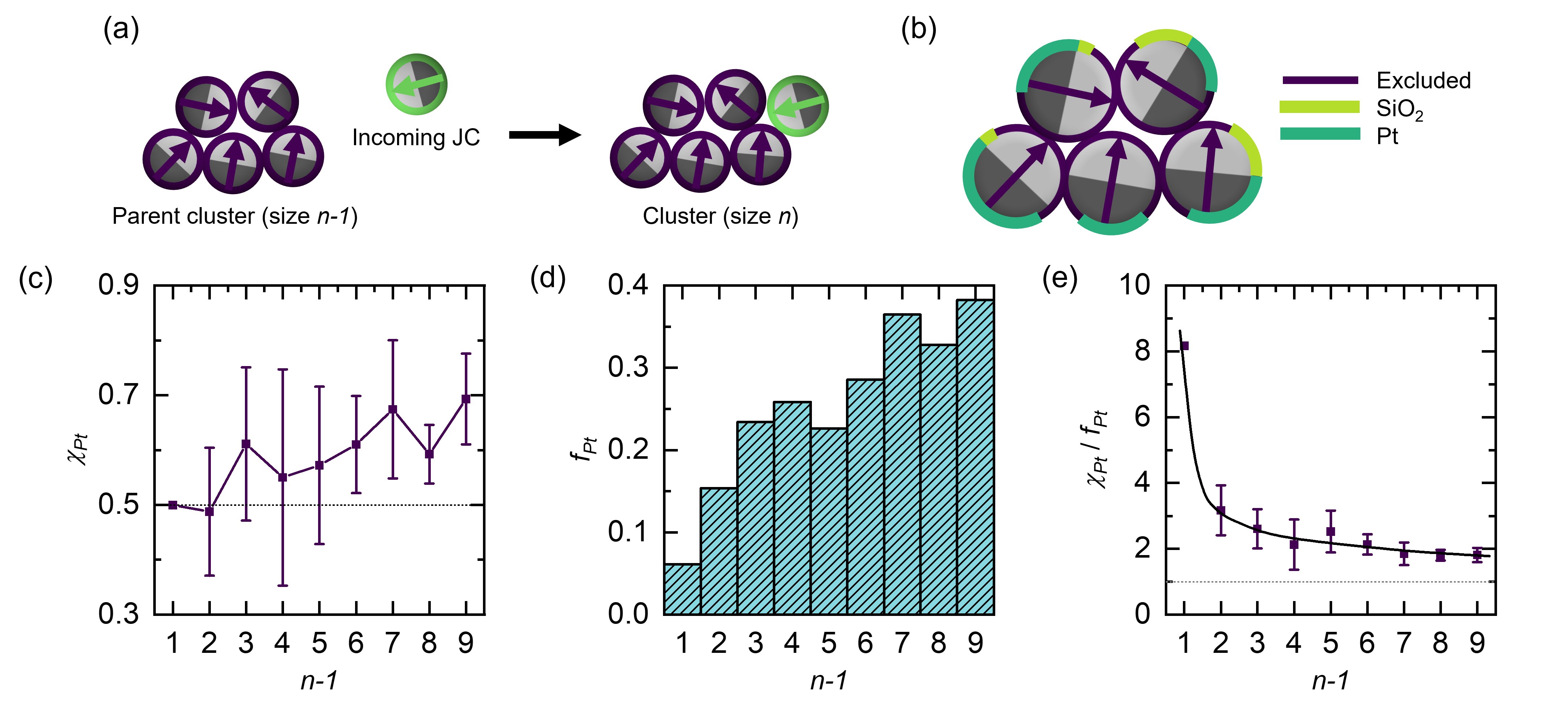}
    \caption{(a) Schematic illustrating the formation of a cluster of size $n$ by single JC contact with an arbitrary $n-1$ sized parent cluster. (b) Schematic of a parent cluster showing the available contactable areas at its perimeter. (c) Variation of the average fraction of the perimeter Pt coverage in the parent cluster. (d) Relative frequency of Pt-side collisions by the incoming particle, at different parent cluster sizes $n-1$. For each data point, a set of at least 75 collisions were considered. (e) Ratio of the  perimeter Pt surface coverage to the frequency of Pt-sided collisions by the incoming JC, as a function of parent cluster size. Solid line is drawn as an eye guide for the reader.}
    \label{fig:Contact}
\end{figure*}

To better understand this loss of alignment, we consider a simple mechanism of cluster formation, illustrated in figure \ref{fig:Contact}(a). Specifically, we examine the process by which a single JC collides with an $(n-1)$-sized parent cluster to form a cluster of size $n$. To understand the effects of chemical interactions in cluster formation, we first examine the surface composition of the parent cluster at its perimeter, and how the single JC interacts with the cluster. While a single isolated JC has equally exposed \ce{SiO2} and Pt surfaces, the parent cluster can have both \ce{SiO2} and Pt surfaces exposed at its perimeter in no specific manner, as they are determined by the cluster shape and the orientations of the constituent JCs. We express the surface availability by $\chi_{Pt}$, the ratio of exposed Pt surface (arc length in 2-D) to the total available surface for incoming JC contact, as shown in figure \ref{fig:Contact}(b). Figure \ref{fig:Contact}(c) shows the variation of $\chi_{Pt}$ across different parent cluster sizes. We find that with increasing cluster size, the Pt coverage at the perimeter gradually rises, reducing the relative availability of \ce{SiO2}, consistent with the previous findings of \citet{Buttinoni2013} where perimeter JCs point inwards the cluster, \emph{i.e.,} exposed coated hemisphere at the cluster perimeter. 

This shift in surface availability has direct implications for how incoming particles attach. Since the \ce{Pt}-side interaction is chemically repulsive, one might expect increasing $\chi_{Pt}$ to reduce successful Pt-side collisions. Yet, figure \ref{fig:Contact}(d) shows that the relative frequency of such Pt-side collisions, $f_{Pt}$, rises with parent cluster size. This apparent contradiction can be explained by the interplay of cluster motility and surface exposure. In the smallest cluster formation, \emph{i.e.,} dimers, collisions mostly occur on the \ce{SiO2}-side or the Janus equator, due to the repulsive chemical field at the \ce{Pt}-side, consistent with the findings from our previous study \cite{Singh2024}. As clusters grow, however, the perimeters become increasingly dominated by exposed \ce{Pt} surfaces, while they slow down. Incoming JCs, being faster and more persistent, thus collide with whichever side they encounter, including the unfavorable Pt-side. This combination of speed mismatch between the cluster and the incoming JC, and the uneven surface exposure in the cluster enables collisions from all directions of the cluster, reducing orientational alignment and hence driving $|\boldsymbol{p}|$ toward zero for larger clusters. This contrasts the interactions between two active JCs of identical size and speed, where collisions in the unfavorable Pt side do not occur \cite{Singh2024}. While the correlation between chemical activity and particle configuration within clusters is intriguing, it falls outside the scope of this study.

Although overall \ce{SiO2}-side collisions remain dominant, the steady rise in $f_{{Pt}}$ suggests a weakening of chemical selectivity in favor of steric constraints. Comparing $\chi_{Pt}$ with $f_{Pt}$, figure \ref{fig:Contact}(e) shows their ratio decreases with size. This trend signals a transition from chemically driven aggregation, biased by surface activity, to a geometry-driven regime dominated by steric hindrance. Thus, cluster growth evolves from anisotropic, guided collisions at small sizes to isotropic, sterically mediated collisions in larger clusters. While similar size-dependent transitions have been reported in diffusioosmotic raft systems \cite{Boniface2024}, our observations arise in freely evolving \ce{SiO2}-Pt colloids through a distinct interplay of motility mismatch and surface exposure.

\section{Conclusions}

In this study, we have experimentally investigated the clustering dynamics of chemically active \ce{SiO2}-Pt JCs. We demonstrated that increasing particle density drives the system into a Dynamic Clustering state marked by continuous formation, breakup, and reorganization of clusters, while maintaining global steady-state statistics. Despite the net repulsive interactions between JCs, the overall clustering dynamics are similar to other chemically active systems found in the literature where chemoattraction drives clustering. 

Beyond establishing the presence of Dynamic Clustering, we investigated how internal cluster structure influences its motion. Our analysis of clusters showed that they exhibit a wide variety of motion, ranging from persistent to cyclic, which depends sensitively on cluster configuration rather than size alone. The introduction of a polarization vector provided a minimal yet powerful descriptor linking internal orientational order to emergent cluster motion, revealing that misalignment among constituent JCs generically leads to circular trajectories. Building on this perspective, we also uncovered a size-dependent crossover in collision mechanisms. While small clusters grow predominantly through chemically selective, orientation-biased encounters, larger clusters experience increasingly isotropic and sterically mediated collisions due to a combination of reduced cluster motility and enhanced exposure of repulsive Pt surfaces at the perimeter. This progressive loss of orientational alignment offers a natural explanation for the observed decrease in cluster polarization with size and reinforces the picture of clustering governed by competing chemical and geometric constraints.

These mechanistic insights also suggest clear routes for controlling cluster behavior. Overall, our findings highlight the potential of active \ce{SiO2}–Pt JCs as a versatile platform for realizing tunable, dynamically reconfigurable clusters with controllable motility and morphology. By establishing a generalized framework that connects cluster structure, internal configuration, and collective dynamics across a broad range of cluster sizes, our study extends prior work largely limited to dimers to assemblies of arbitrary size. This framework opens avenues for systematically engineering cluster behavior through controlled modifications. For example, rendering the \ce{SiO2} hemisphere hydrophobic \cite{Gao2013} to enhance \ce{SiO2}-\ce{SiO2} contacts and assemble stable cyclic clusters capable of generating localized fluid flows, with potential relevance for micromixing applications. More broadly, introducing controlled heterogeneity, such as passive tracers, mixtures of JCs with different sizes or compositions, or embedding the system in polymeric or viscoelastic media offers promising directions for uncovering new dynamical regimes and transitions. These results lay the groundwork for leveraging active Janus colloid clusters in applications requiring programmable collective motion and adaptive organization, ranging from micro-robotic actuation to the design of responsive and functional soft materials.

\section*{Supplementary Material}
See the Supplementary Material for Supplementary movies (S1–S5), Orientation Autocorrelation (OACF) and Mean Square Displacement (MSD) curves of isolated active Janus Colloids, Variation of the average instantaneous speed with area fraction, Variation of the cluster size retention time with cluster size, and OACF and MSD curves of a few representative dimers at different orbit radii.

\begin{acknowledgments}
The authors acknowledge the funding received from the Science and Engineering Research Board (Grant Nos. ECR/2018/000401) and from the Department of Science and Technology, India (Grant No. SR/FST/ETII-055/2013).
\end{acknowledgments}

\section*{Author Declarations}
\subsection*{Conflicts of Interest}
The authors have no conflicts to disclose.

\subsection*{Author Contributions}
R.M. led the conceptualization of the study and supervised the project. H.R. performed the experiments, and led the formal analysis and visualization, with contributions from A.S., A.K., M.P., and P.K. H.R., A.C., and R.M. wrote the paper. All authors reviewed the manuscript.

\section*{Data Availability Statement}
The data that support the findings of this study are available from the corresponding author upon reasonable request.

\section*{References}
\bibliography{References}% Produces the bibliography via BibTeX.

\clearpage
\onecolumngrid          % if using revtex
\setcounter{figure}{0}

\renewcommand{\thefigure}{S\arabic{figure}}

\section*{Supporting Figures}

\begin{figure}[h]
    \centering
    \includegraphics[width=\textwidth]{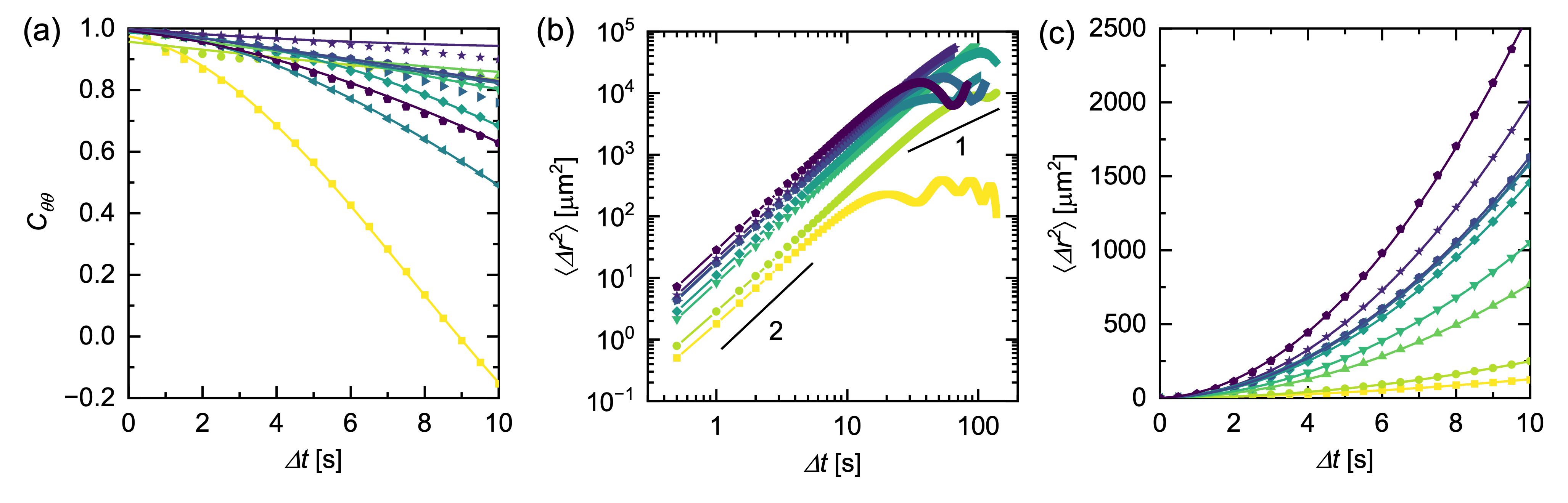}
    \caption{\textbf{Dynamics of Isolated \ce{SiO2}-Pt Janus Colloids (JCs)}: (a) Fitted Orientation Autocorrelation Function (OACF) curves for a few representative isolated \ce{SiO2}-Pt JCs. (b) Corresponding Mean Square Displacement (MSD) curves. (c) Fitted MSD curves at shorter $\Delta{t}$.}
    \label{fig:enter-label}
\end{figure}

\begin{figure}[h]
    \centering
    \includegraphics[width=\textwidth]{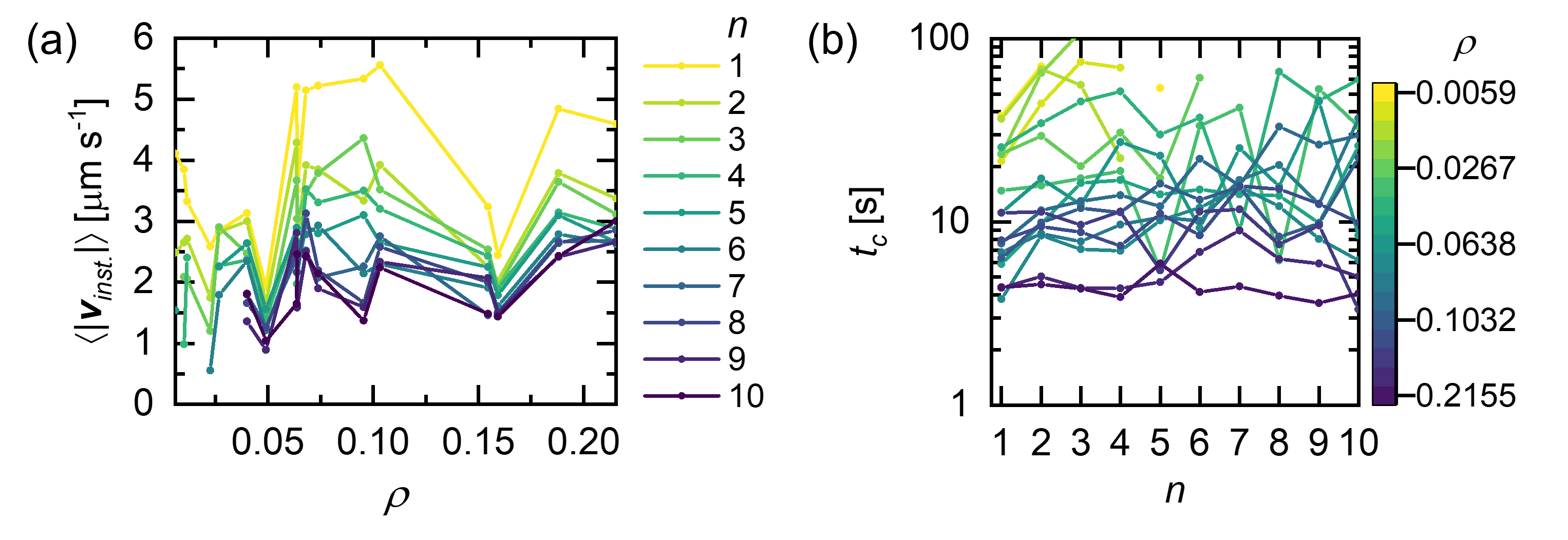}
    \caption{(a) Variation the average instantaneous speed $\langle|\textbf{v}_{inst.}|\rangle$ (mean) with area fraction $\rho$ for upto 10-particle clusters, and (b) Variation of the cluster size retention time $t_c$ (mean) with size $n$ at various area fractions $\rho$. Both $\langle|\textbf{v}_{inst.}|\rangle$ and $t_c$ show no systematic variation with $\rho$ and $n$ respectively.  Error bars have been excluded in both figures for visualization clarity.}
    \label{fig:enter-label}
\end{figure}

\begin{figure}[h]
    \centering
    \includegraphics[width=\textwidth]{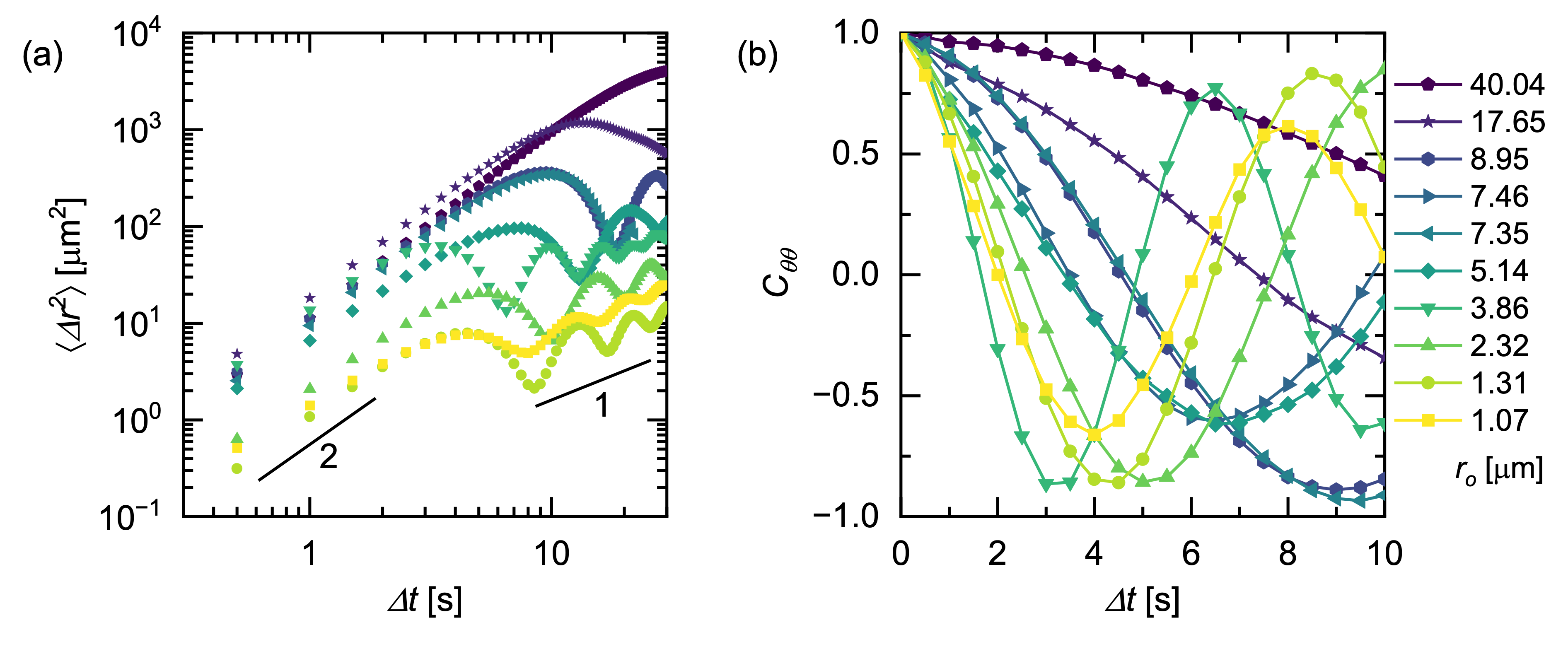}
    \caption{(a) MSD curves for a few representative dimers covering a wide range of orbit radii $r_o$. (b) Corresponding OACF curves.}
    \label{fig:enter-label2}
\end{figure}

\end{document}